\documentclass[11pt,twoside]{article}

\usepackage{asp2006}
\usepackage[dvips]{graphicx}

\begin{document}
\title{Surprising Evolution of Faraday Rotation Gradients in the Jet of 1803+784}

\author{Mehreen Mahmud, Denise Gabuzda}
\affil{Department of Physics, University College Cork, Cork, Ireland}

\author{Vladislavs Bezrukovs}
\affil{Department of Physics, Cork Institute of Technology, Cork, Ireland}

\begin{abstract} 

Several multi-frequency polarization studies have shown the presence of systematic Faraday Rotation gradients across the parsec-scale jets of Active Galactic Nuclei (AGN), taken to be due to the systematic variation of the line-of-sight component of a helical magnetic field across the jet. Other studies have confirmed the presence and sense of these gradients in several sources, thus providing evidence that these gradients persist over time and over large distances from the core. However, we find surprising new evidence for a reversal in the direction of the Faraday Rotation gradient across the jet of 1803+784, for which multi-wavelength polarization observations are available at four epochs. At all four epochs, we observe transverse Rotation Measure (RM) gradients across the jet, consistent with the presence of a helical magnetic field wrapped around the jet. However, we also observe a ``flip'' in the direction of the gradient between June 2000 and August 2002. Although the origins of this phenomena are not understood, one way to interpret this change is if the sense of rotation of the central supermassive black hole and accretion disc has remained the same, but the dominant magnetic pole facing the Earth has changed from North to South. 

\end{abstract}

\section{Introduction}   

1803+784 has been studied using VLBI for more than two decades. The predominant jet direction in centimeter-wavelength images is toward the West, and the dominant jet magnetic (B) field is perpendicular to the local jet direction throughout the jet (Gabuzda \& Chernetskii 2003, Veres \& Gabuzda, these proceedings). It seems likely that this transverse jet B field represents the ordered toroidal component of the intrinsic jet B field. 

Faraday Rotation of the plane of linear  polarization occurs during the passage of an electromagnetic wave through a region with free electrons and a magnetic field with a non-zero component along the line-of-sight, and is due to the difference in the propagation velocities of the right and left-circularly polarized components of the wave. The amount of rotation is proportional to the density of free electrons $n_{e}$ multiplied by the line-of-sight magnetic field $B \cdot dl$, the square of the observing wavelength $\lambda^{2}$, and various physical constants; the coefficient of $\lambda^{2}$  is called the Rotation Measure, RM:

\begin{eqnarray}
           \Delta\chi\propto\lambda^{2}\int n_{e} B\cdot dl\equiv RM\lambda^{2}
\end{eqnarray}

Systematic gradients in the Faraday Rotation have been observed across the parsec-scale jets of several AGN, interpreted as reflecting the systematic change in the line-of-sight component of a toroidal or helical jet B field across the jet; such fields would come about in a natural way as a result of the ``winding up'' of an initial ``seed'' field by the rotation of the central accreting objects (Asada et al 2002; Gabuzda et al 2004; Mahmud \& Gabuzda, these proceedings).

\section{Observations and Reduction}
Polarization data for 1803+784 were obtained using the 10 25-m radio telescopes of the Very Long Baseline Array (VLBA) at four different epochs: 6 April 1997, 27 June 2000 (Zavala \& Taylor 2003), 24 August 2002 and 22 August 2003. The observations for 1997 were obtained at 5GHz, 8GHz, 15GHz and 22GHz; for 2000 at 7 frequencies ranging from 8.1 and 15.2GHz (Zavala \& Taylor 2003); for 2002 at 15GHz, 22GHz and 43GHz; and for 2003 at 6 frequencies ranging from 4.6GHz and 15.4GHz. The observations for this last epoch were obtained as part of a study of about three dozen BL Lac objects; see Mahmud \& Gabuzda, these proceedings. We calibrated and imaged the data for 1997, 2002 and 2003 in the NRAO AIPS package using standard techniques and analyzed the results together with the RM map for June 2000 published by Zavala and Taylor (2003).

After matching the imaging parameters and beam sizes of the final images at  all the wavelengths, we constructed maps of the RM,  after first subtracting the effect of the integrated RM (presumed to arise in our Galaxy) from the observed polarization angles (Pushkarev 2001).


 \begin{figure}[!ht]
  \begin{minipage}[t]{13.5cm}
  \begin{center}
 \includegraphics[width=11.0 cm,clip]{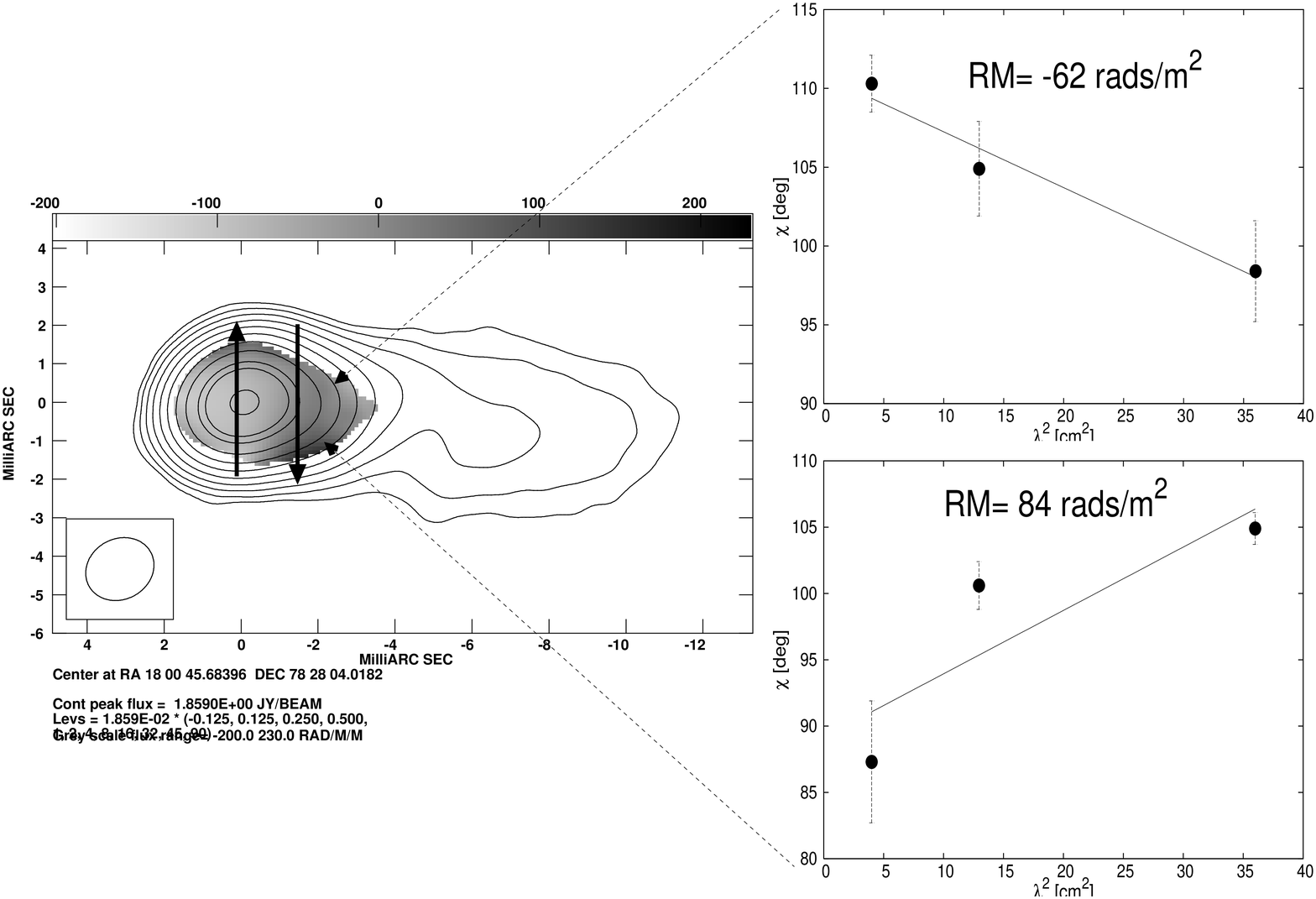}
  \caption{RM map of 1803+784 observed on 6 April 1997, based on observations at 15, 8.4 and 5GHz with the I contour map at 8.4GHz overlaid. The errors shown are 2 $\sigma$.
}
\end{center}
  \end{minipage}
  \hfill

  \begin{minipage}[t]{13.5cm}
  \begin{center}
  \includegraphics[width=11.0 cm,clip]{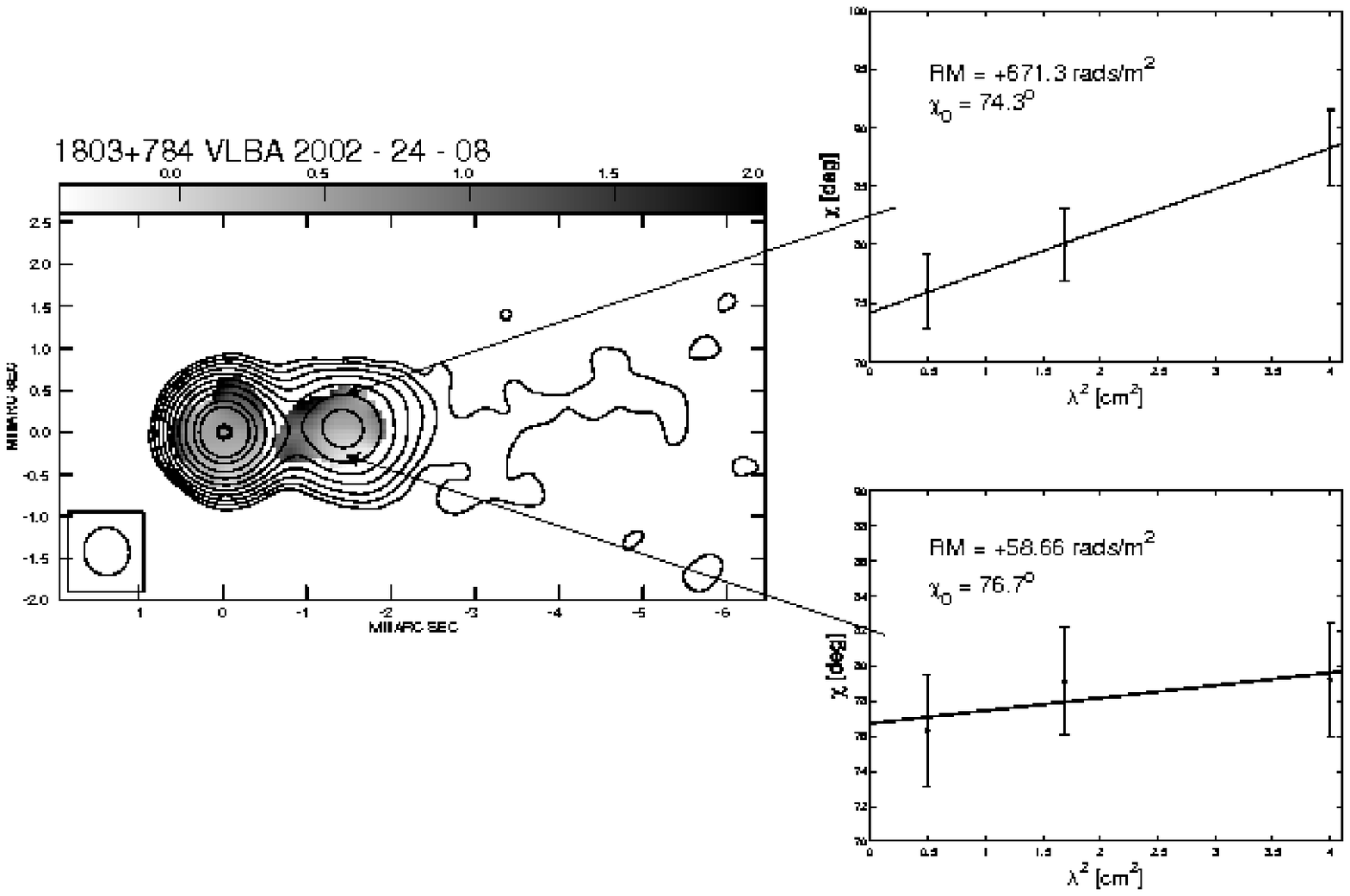}
 \caption{RM Map of 1803+784 observed on 24 August 2002, based on observations at 43, 22 and 15GHz with the I contours at 15GHz overlaid. The errors shown in the plots are 2 $\sigma$.}
  \end{center}
  \end{minipage}
 \end{figure}


\begin{figure}[!hp]
 \begin{center}
 \includegraphics[width=10.0 cm, clip]{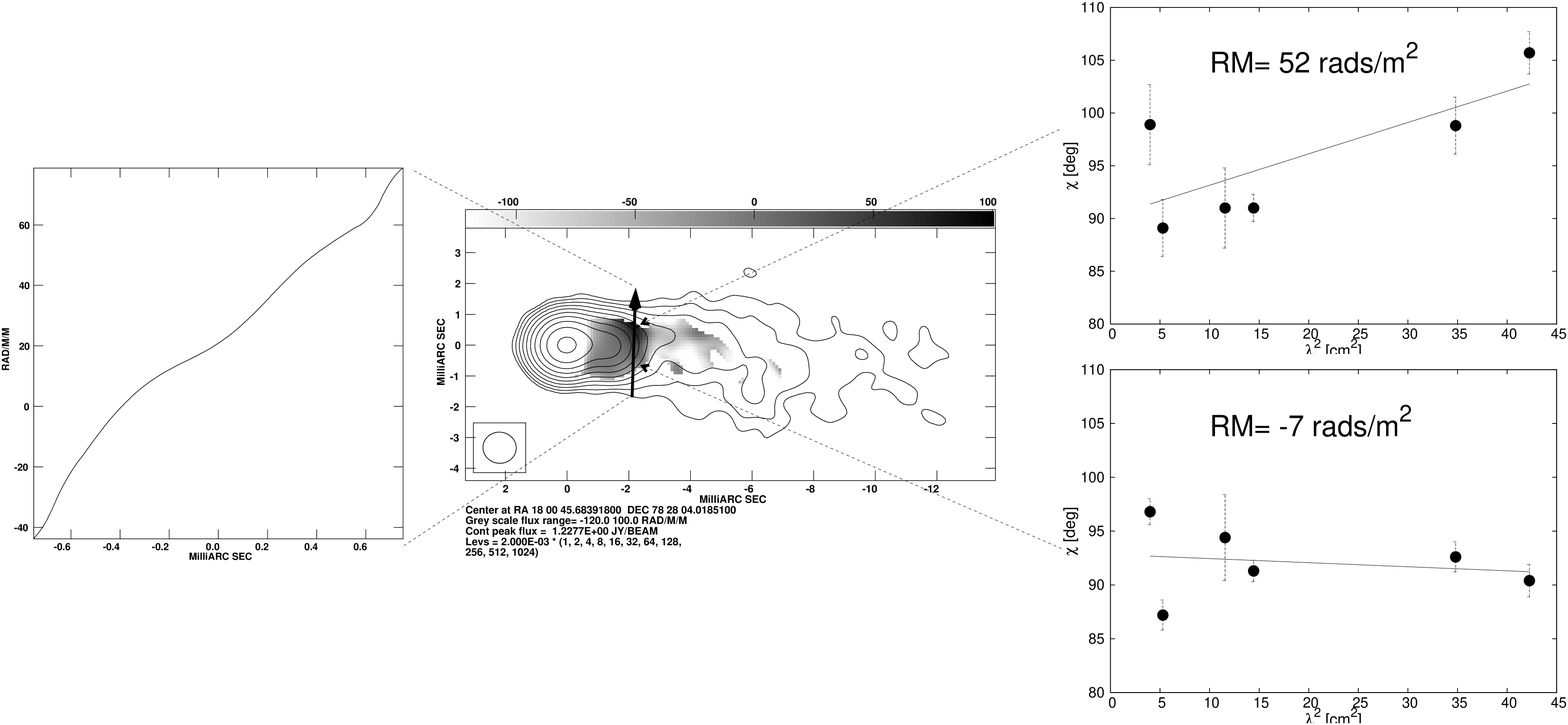}
  \caption{RM map for the jet of 1803+784, observed on 22 August 2003, at six frequencies from 15 to 5GHz with the I contours at 8.9GHz overlaid. The errors in the plots are 1 $\sigma$ . Also presented are RM slices across the jet where we observe the gradient. The gradient is clearly visible in these cross-sections.}
\end{center}
 \end{figure}
\begin{figure}[!hp]
  \begin{center}
  \includegraphics [width=10.0 cm, clip]{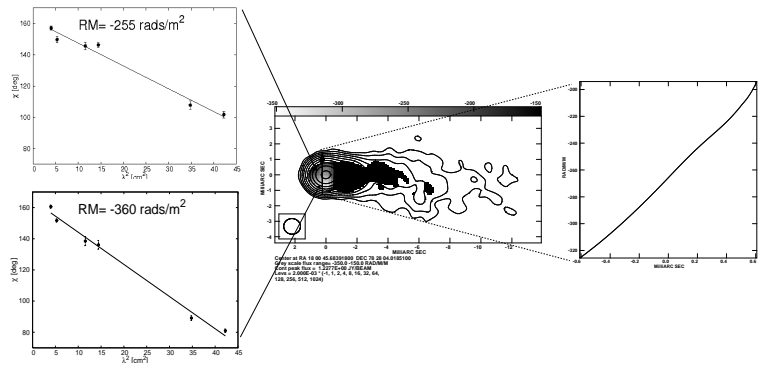}
\caption{RM map for the core region of 1803+784 (epoch 22 August 2003) with the I contours at 8.9GHz overlaid. The errors in the plots are 1 $\sigma$ . Also presented are RM slices across the core region where we observe the gradient. The gradient is clearly visible in these cross-sections.}
  \end{center}
  \end{figure}
  \begin{figure}[!hp]
  \begin{center}
  \includegraphics[width= 6.5 cm]{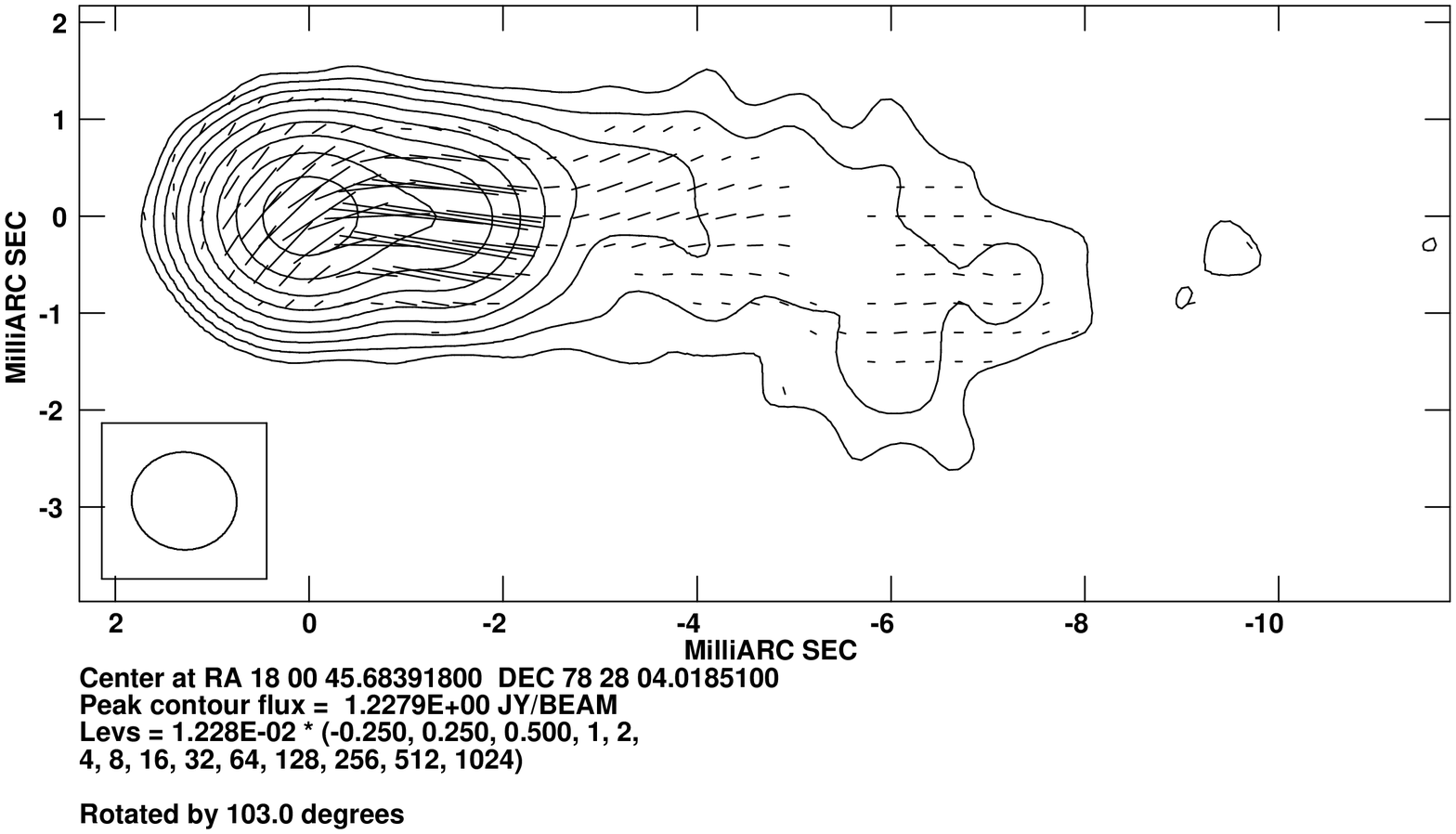}
\caption{Stokes I map of 1803+784 (epoch 22 August 2003) at 7.9GHz, showing the Electric Field  vectors, roughly aligned with the jet. The magnetic (B) field vectors are thus perpendicular to the jet direction.}
  \end{center}
 
 \end{figure}


\begin{figure}[ht]
 \centering
 \includegraphics[width=8.0 cm, clip]{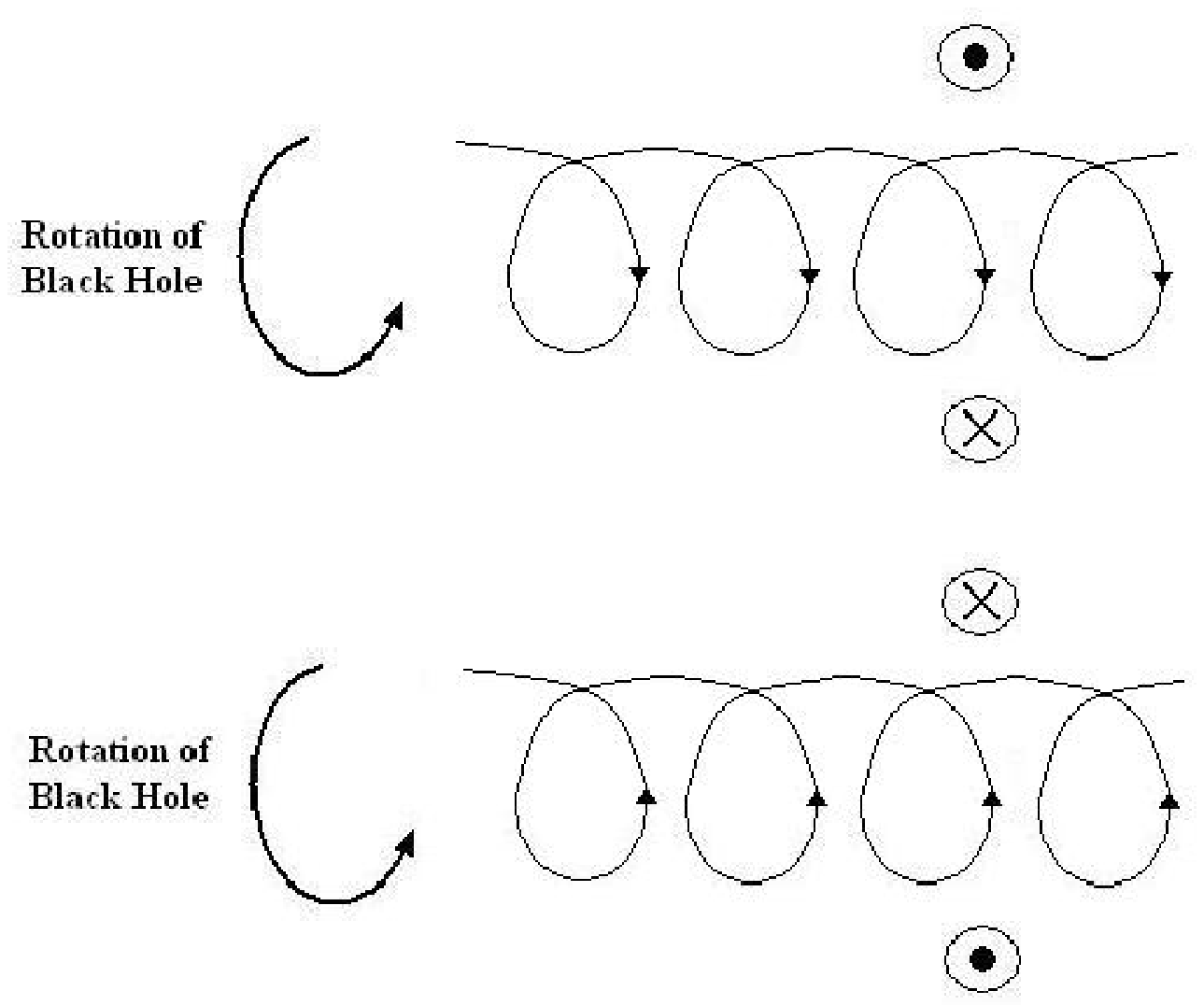}
 \caption{Illustration of how a ``flip'' in the RM gradient can be caused by a change in the polarity of the black-hole B-field facing the Earth. The top diagram corresponds to a N pole, and the bottom diagram to a S pole. The arrows show the direction of the B field, and are marked on the nearer side of the helical field to the observer.}
\end{figure}

\section{Results and Discussion}
The images in Figures 1-4 show total intensity contours with RM distributions superposed for each of our three epochs. The arrows show the direction of the RM gradients in the corresponding regions. The accompanying panels show plots of polarization angle ($\chi$) vs. wavelength ($\lambda$) squared for the indicated region. The uncertainties in the polarization angles shown in the plots were calculated using the rms noise in the Stokes Q and U maps, and also include a contribution for the uncertainty in the polarization angle calibration.

Transverse RM gradients are visible across the VLBI jet of 1803+784 at all three of our epochs, on scales of $\sim$ 2-3 mas (Figs. 1-3); we are tentatively able to follow it even further from the core. This strongly supports the hypothesis that this jet has a helical magnetic field, consistent with the observed transverse magnetic-field structure (Fig. 5). 

The RM map of Zavala \& Taylor (2003) also shows a clear transverse gradient, with a negative RM along the Northern edge of the jet and less negative or positive RM along the Southern edge (Similar to the RM map in Fig. 1) An important feature that has emerged is the reversal in the RM gradient in the jet of 1803+784 between June 2000 (Zavala \& Taylor 2003) and Aug 2002 (Fig. 2). The RM values increases toward the Southern edge of the jet in the April 1997 and June 2000 RM maps but toward the Northern edge of the jet in the August 2002 and August 2003 RM maps. The origin for this ``flip'' in the direction of the transverse RM gradient in the jet of 1803+784 is not clear. No radical change in the intensity or polarization structure of the VLBI jet accompanied the observed RM-gradient flip. One possible way to retain a transverse RM gradient in a helical magnetic field model but reverse the direction of this gradient is if the direction of rotation of the central black hole (i.e.~the direction in which the field threading the accretion disc is ``wound up'') remains constant, but the ``pole'' of the black hole facing the Earth changes from North to South, or vice versa (Fig. 6). The origin of this phenomena is likewise not clear, but it is reminiscent of the reversal of the magnetic polarity observed for the Sun during the solar activity cycle, associated with the dynamo mechanism. 
We also observe gradients in the core for all 3 of our epochs, all of which are in the same direction (i.e., the ``flip'' of the gradient occurred only in the jet region). Further multi-wavelength polarization studies of this source are clearly crucial for our understanding of how these Faraday Rotation gradients evolve over time and with distance from the core.

\acknowledgements 
This publication has emanated from research conducted with the financial support of Science Foundation Ireland. The National Radio Astronomy Observatory is operated by Associated Universities Inc.



\begin{thebibliography}{}

 \bibitem{} Asada, K., Inoue, M., Uchida, Y., Kameno, S., Fujisawa, K., Iguchi, S. Mutoh, M. 2002, PASJ, 54, L39
\bibitem[]{}Gabuzda, D. C., Chernetskii, V. A. 2003, MNRAS, 339, 3, 669-679
\bibitem{}{}Gabuzda, D. C., Murray, E. \& Cronin, P. 2004, MNRAS, 351, L89 - L93 
\bibitem[]{}Pushkarev, A. 2001, Astron. Rep., 45, 667
\bibitem[]{}Rusk, R. 1988, PhD Thesis, University of Toronto
\bibitem[]{}Zavala, R. \& Taylor, G. 2003, ApJ, 589, 126

\end{thebibliography}
\end{document}